\documentclass[prl,floatfix,showpacs,twocolumn,superscriptaddress]{revtex4-1}

\newcommand{\Eq}[1]{Eq.~(\ref{eq:#1})}
\newcommand{\Eqs}[2]{Eqs.~(\ref{eq:#1}) and (\ref{eq:#2})}
\newcommand{\Fig}[1]{Fig.~\ref{fig:#1}}
\newcommand{\Figs}[1]{Figs.~\ref{fig:#1}}

\newcommand{\Ref}[1]{Ref.~\cite{#1}}
\newcommand{\rr}{\mathbf{r}}

\newcommand{\f}{\mathbf{f}}

\renewcommand{\v}{\mathbf{v}}

\newcommand{\gdot}{\dot\gamma}
\newcommand{\RD}{RD$_0$}
\newcommand{\CD}{CD$_0$}
\newcommand{\phieff}{\phi_\mathrm{eff}}

\newcommand{\skippa}[1]{}

\newlength{\figwidth}
\usepackage{graphicx}

\begin{document}
\title{Universality of Jamming Criticality in Overdamped Shear-Driven Frictionless Disks}

\author{Daniel V{\aa}gberg}
\author{Peter Olsson}
\affiliation{Department of Physics, Ume\aa\ University, 901 87 Ume\aa, Sweden}
\author{S. Teitel}
\affiliation{Department of Physics and Astronomy, University of Rochester,
  Rochester, NY 14627}
\date{\today}   
\begin{abstract}
  We investigate the criticality of the jamming transition for overdamped shear-driven
  frictionless disks in two dimensions for two different models of energy dissipation: (i)
  Durian's bubble model with dissipation proportional to the velocity difference of
  particles in contact, and (ii) Durian's ``mean-field'' approximation to (i), with
  dissipation due to the velocity difference between the particle and the average uniform
  shear flow velocity.  By considering the finite-size behavior of pressure, the pressure
  analog of viscosity, and the macroscopic friction $\sigma/p$, we argue that these two
  models share the same critical behavior.
\end{abstract}
\maketitle

Many different physical systems, such as granular materials, suspensions, foams and
emulsions, may be modeled in terms of particles with short ranged repulsive contact
interactions.  As the packing fraction $\phi$ of such particles is increased, the system
undergoes a jamming transition from a liquid state to a rigid but disordered solid.  It
has been proposed that this jamming transition is a manifestation of an underlying
critical point, ``point J'', with associated scaling properties such as is found in
equilibrium phase transitions \cite{Liu_Nagel, Liu_Nagel_vanSaarloos_Wyart}.  Scaling
properties are indeed found when such systems are isotropically compressed, with pressure,
elastic moduli, and contact number increasing as power laws as $\phi$ increases above the
jamming $\phi_J$ \cite{OHern_Silbert_Liu_Nagel:2003}.  When such systems are sheared with
a uniform strain rate $\dot\gamma$, a unified critical scaling theory has successfully
described both the vanishing of the yield stress as $\phi\to\phi_J$ from above, the
divergence of the shear viscosity as $\phi\to\phi_J$ from below, and the non-linear
rheology exactly at $\phi=\phi_J$ \cite{Olsson_Teitel:jamming}.

One of the hallmarks of equilibrium critical points is the notion of \emph{universality};
the critical behavior, specifically the exponents describing the divergence or vanishing
of observables, depend only on the symmetry and dimensionality of the system, and not on
details of the specific interactions.  For statically jammed states created by
compression, where only the elastic contact interaction comes into play, it is understood
that the relevant critical exponents are simply related to the form of the elastic
interaction, and are all simple rational fractions \cite{OHern_Silbert_Liu_Nagel:2003}.  In
contrast, shear-driven steady states are formed by a balance of elastic and dissipative
forces, and it is thus an important question whether or not the specific form taken for
the dissipation is crucial for determining the critical behavior.

In a recent work by Tighe et al.\ \cite{Tighe_WRvSvH}, it was claimed that changing the
form of the dissipation can indeed alter the nature of the criticality for sheared
overdamped frictionless disks.  In contrast to earlier work \cite{Olsson_Teitel:jamming},
where particle dissipation was taken with respect to a uniformly sheared background
reservoir, Tighe et al.\ used a collisional model for dissipation.  They argued that this
change in dissipation resulted in dramatically different behavior from that found
previously, specifically (i) there is no length scale $\xi$ that diverges upon approaching
$\phi_J$, and so behavior can be described analytically with a mean-field type model; (ii)
critical exponents are simple rational fractions; (iii) there is no single critical
scaling, but rather several different flow regimes, each with a different scaling.  In
this work we numerically re-investigate the model of Tighe et al.\ and present results
arguing against these conclusions.  In particular we conclude that the two models have
rheology that is characterized by the \emph{same} critical exponents, and so are in the
same critical universality class.

We simulate bidisperse frictionless disks in two dimensions (2D), with equal numbers of
big and small disks with diameter ratio 1.4, at zero temperature. The interaction of disks
$i$ and $j$ in contact is $V_{ij} = k_e \delta_{ij}^2/2$, where the overlap is
$\delta_{ij}=r_{ij}/d_{ij}-1$, with $d_{ij}$ the sum of the disks' radii.  The elastic
force on disk $i$ is $\f^\mathrm{el}_i = -\nabla_i \sum_jV_{ij}$, where the sum is over
all particles $j$ in contact with $i$. We use Lees-Edwards boundary conditions 
\cite{Evans_Morriss} to introduce a time-dependent uniform shear strain $\gamma(t) = \gdot
t$ in the $\hat x$ direction.

We consider two different models for energy dissipation.  The first, which we call
``contact dissipation'' (CD), is the model introduced by Durian for bubble dynamics in
foams \cite{Durian:1995}, and is the model used by Tighe et al.\ \cite{Tighe_WRvSvH}.  Here
dissipation occurs due to velocity differences of disks in contact,
\begin{equation}
  \f^\mathrm{dis}_{\mathrm{CD},i} = -k_d \sum_j (\v_i - \v_j),\qquad\v_i=\dot\mathbf{r}_i.
\end{equation}
In the second, which we call ``reservoir dissipation'' (RD), dissipation is with respect to
the average shear flow of a background reservoir,
\begin{equation}
  \f^\mathrm{dis}_{\mathrm{RD},i} = -k_d (\v_i - \v_\mathrm{R}(\rr_i)),\qquad
  \v_\mathrm{R}(\rr_i)\equiv\gdot y_i\hat x.
\end{equation}
RD was also introduced by Durian \cite{Durian:1995} as a ``mean-field'' \cite{Tewari:1999}
approximation to CD, and is the model used in many earlier works on criticality in shear driven
jamming \cite{Tewari:1999, Olsson_Teitel:jamming, Andreotti:2012, Lerner-PNAS:2012}.

The equation of motion for both models is
\begin{equation}
  m_i\dot\v_i = \f^\mathrm{el}_i +\f^\mathrm{dis}_i.
\end{equation}
Here we are interested in the overdamped limit, $m_i\to0$ \cite{Durian:1995}.  In RD it is
straightforward to set $m_i=0$, in which case the equation of motion becomes simply $\v_i
= \v_\mathrm{R}(\rr_i) +\f_i^\mathrm{el}/k_d$; we call this limit RD$_0$. In CD, because
the dissipation couples velocities one to another, setting $m_i=0$ effectively requires
inverting the matrix of contacts to rewrite the equation of motion in a form suitable for
numerical integration.  Instead of that numerically difficult procedure, our approach here
is to simulate particles with a finite mass, and verify that the mass is small enough for
the system to be in the overdamped $m_i\to 0$ limit; we call this limit CD$_0$.  For our
simulations we use units in which $k_e=k_d=1$, length is in units such that the small disk
diameter $d_s=1$, time in units of $\tau_0 \equiv k_d d_s^2/k_e=1$, and particles of equal
mass density, such that $m_i$ for a particle of diameter $d_i$ is $m_i=2m\pi d_i^2/4$,
with $m=1$.  In our Supplemental Material \cite{suppMat} we confirm that this choice is
sufficient to be in the $m_i\to 0$ limit.  For \RD\ our simulations use $N=65536$
particles, while for \CD\ we use $N=262144$, unless otherwise noted.  For \RD\ our slowest
strain rate is $\gdot=10^{-9}$, while for \CD\ we can reach only $\gdot=10^{-7}$.

Before presenting our evidence that the two models \RD\ and \CD\ have the same critical
rheology, we first comment on one quantity that is clearly very different in the two
models, the transverse velocity correlation, $g_y(x)\equiv\langle v_y(0)v_y(x)\rangle$.
In \RD\, $g_y(x)$ shows a clear minimum at a distance $x=\xi$, and this length diverges as
one approaches the critical point $(\phi_J, \gdot\to0)$ \cite{Olsson_Teitel:jamming}.  In
\CD\, however, it was found \cite{Tighe_WRvSvH} that $g_y(x)$ decreases monotonically
without any obvious strong dependence on either $\phi$ or $\gdot$.  This led Tighe et al.\
to conclude that there is no diverging length $\xi$ in model CD$_0$, that the only
macroscopic length scale is the system length $L$, and thus there are no critical
fluctuations.  In our own work we have confirmed this dramatic difference in the behavior
of $g_y(x)$, but see our Supplemental Material for further comments \cite{suppMat}.

However the apparent absence of a diverging $\xi$ in $g_y(x)$ for \CD\ does not
necessarily imply that such a diverging length does not exist.  In the following we
present evidence for such a diverging $\xi$ in \CD\ by considering the finite-size
dependence of the pressure $p$ as a function of strain rate $\gdot$ at $\phi_J$.  By a
critical scaling analysis of the pressure analogue of viscosity, $\eta_p\equiv p/\gdot$,
we further show that the rheology in \CD\ is characterized by the same critical exponents
as is RD$_0$.  Finally we consider the macroscopic friction $\mu\equiv \sigma/p$ in the
two models, with $\sigma$ the shear stress, and show that they behave similarly.  There is
no sign of the roughly square root vanishing of $\mu$ at $\phi_J$ that would be expected from the
model of Tighe et al., but rather $\mu$ appears to be finite passing through $\phi_J$, as
found in recent experiments on foams \cite{Lespiat}.

\emph{Finite-size dependence of pressure}: 
We consider here only the elastic part of $p$ which is computed from the elastic
contact forces in the usual way \cite{OHern_Silbert_Liu_Nagel:2003}.  If jamming behaves
like a critical point, we expect $p$ to obey finite-size-scaling at large system lengths
$L$ \cite{Olsson_Teitel:gdot-scale},
\begin{equation}
  p(\phi,\gdot,L)=L^{-y/\nu}{\cal P}((\phi-\phi_J)L^{1/\nu},\gdot L^{z}).
\label{epL0}
\end{equation}
Exactly at $\phi=\phi_J$ the above becomes \cite{Hatano:2011}
\begin{equation}
  p(\phi_J,\gdot,L)=L^{-y/\nu}{\cal P}(0,\gdot L^z).
\label{epL}
\end{equation}
For sufficiently small $L$, where $\gdot L^z\ll 1$, we get $p\sim L^{-y/\nu}$.  For
sufficiently large $L$, where $\gdot L^z\gg 1$, $p$ becomes independent of $L$ and so
$p\sim \gdot ^{y/z\nu}$.  The crossover occurs when $L=\xi$ at $\gdot L^z\approx
1\Rightarrow \xi\sim \gdot^{-1/z}$, giving a diverging correlation length as $\gdot\to
0$.

\begin{figure}
  \includegraphics[width=3.5in]{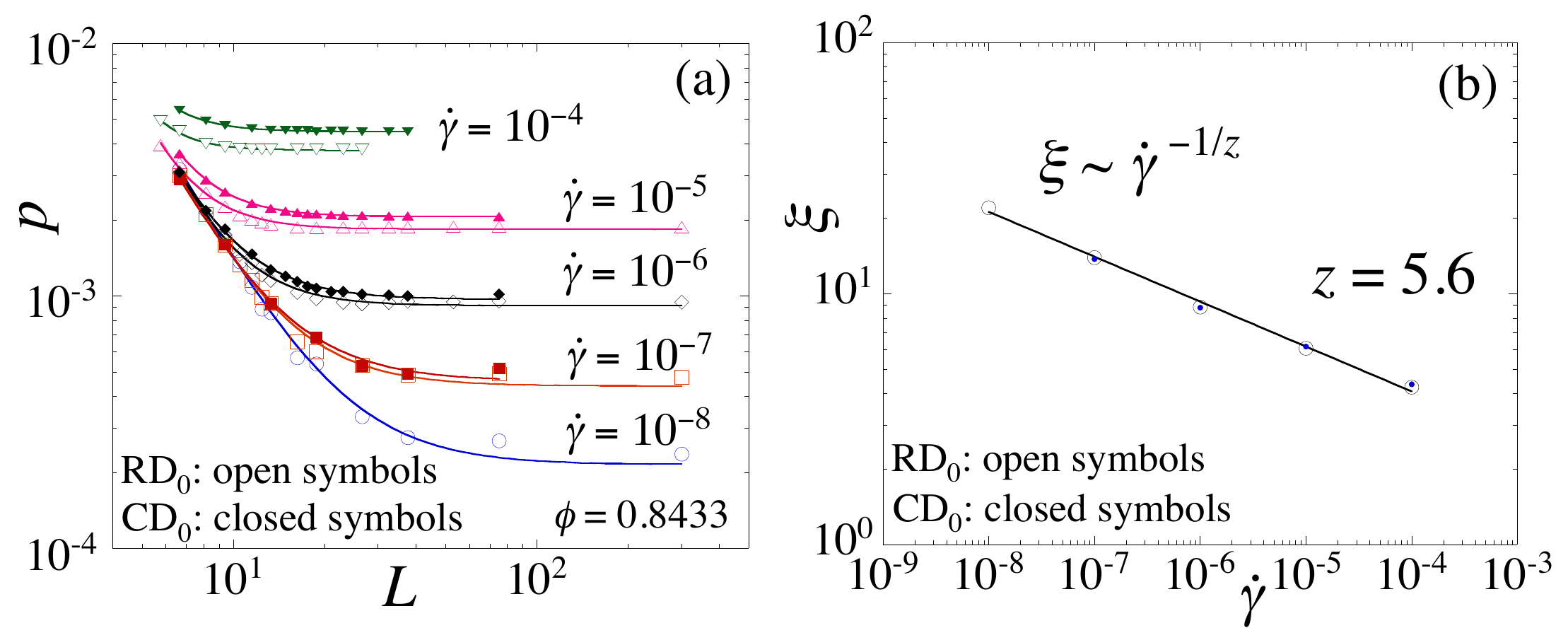}
  \caption{(a) Finite size behavior of pressure $p$ in model \RD\ (open symbols) and model
    CD$_0$ (closed symbols) at different strain rates $\gdot$. For $\gdot = 10^{-8}$ we
    only have results for model RD$_0$. The crossover from power law behavior at small $L$
    to a constant at large $L$ determines the correlation length $\xi$, plotted vs $\gdot$
    in (b).  We see that $\xi$ is essentially identical in both models, growing
    monotonically as $\gdot$ decreases, reaching values as large as $\xi \approx 20$ for
    our smallest $\gdot$. As $L$ varies, the number of particles varies from $N=24$ to $4096$.}
  \label{fig:fss-p}
\end{figure}

In \Fig{fss-p}(a) we plot $p$ vs.\ $L$ for \RD\ and \CD\ at $\phi=0.8433\approx\phi_J$ for
several different $\gdot$. Both models clearly behave similarly.  To determine the
crossover $\xi$ we fit our data to the simple empirical form $p=C(1+[\xi/L]^x)$ that
interpolates between the two asymptotic limits. This fit gives the solid lines in
\Fig{fss-p}(a). The resulting $\xi$ is plotted in \Fig{fss-p}(b).  We see that $\xi$ is
essentially identical in the two models, growing monotonically as $\gdot$ decreases,
reaching values as large as $\xi\simeq 20$ for our smallest $\gdot$.  Such a large length,
many times the microscopic length set by the particle size, is clear evidence for
cooperative behavior \cite{glassComment}. Thus our results indicate a growing macroscopic
length scale in CD$_0$, just as was found for RD$_0$. The exponent $z\approx 5.6$ found in
Fig.~\ref{fig:fss-p}(b) must, however, be viewed with caution since corrections-to-scaling
are large at the sizes $L$ considered here \cite{Vagberg_VMOT:jam-fss}, and the neglect of
such corrections can skew the resulting effective exponents away from their true values at
criticality.  See our Supplemental Material \cite{suppMat} for a more in depth discussion.

\emph{Pressure analog of viscosity}: We now seek to compute the critical exponents of the
two models \RD\ and CD$_0$, to see if they are indeed in the same universality class.  To
do this we consider data at various packing fractions $\phi$ and strain rates $\gdot$
close to the jamming transition.  We use system sizes large enough ($N=65536$ for \RD\ and
$N=262144$ for CD$_0$), such that finite size effects are negligible for the data
presented here.  As in our recent work on \RD\ \cite{Olsson_Teitel:gdot-scale}, we
consider here the pressure analog of viscosity $\eta_p\equiv p/\gdot$, since corrections
to scaling are smaller for $p$ than for shear stress
$\sigma$ \cite{Olsson_Teitel:gdot-scale}.

To extract the jamming fraction $\phi_J$ and the critical exponents we use a mapping from
our system of soft-core disks to an effective system of hard-core disks.  We have
previously shown \cite{Olsson_Teitel:jam-HB} this approach to give excellent agreement
with results from a more detailed two variable critical scaling analysis 
\cite{Olsson_Teitel:gdot-scale} for RD$_0$.  We use it here because it requires no
parameterization of an unknown crossover scaling function and so is better suited
particularly to \CD\ where the range of our data is more limited ($10^{-7}\le \gdot$) as
compared to \RD\ ($10^{-9}\le\gdot$).

This method assumes that the soft-core disks at $\phi$ and $\gdot$ can be described as
effective hard-core disks at $\phi_\mathrm{eff}(\gdot)$, by modeling overlaps as an
effective reduction in particle radius \cite{Olsson_Teitel:jam-HB}. Measuring the overlap
via the average energy per particle $E$, we take
\begin{equation}
  \phieff = \phi-cE^{1/2y},
  \label{eq:phieff}
\end{equation}
where $y$ is the exponent with which the pressure rises as $\phi$ increases above $\phi_J$
along the yield stress curve $\gdot\to 0$, as in Eq.~(\ref{epL0}), and $c$ is a constant.
We can then express the viscosity of this effective hard-core system as,
\begin{equation}
  \eta_p(\phi,\gdot) = \eta_p^\mathrm{hd}(\phieff) 
  = A (\phi_J - \phieff)^{-\beta}.
  \label{eq:etaphd}
\end{equation}
Our analysis then consists of adjusting $\phi_J$, the exponents $y$ and $\beta$, and the
constants $c$ and $A$ in \Eqs{phieff}{etaphd}, to get the best possible fit to our data.

\begin{figure}
  \includegraphics[width=1.65in]{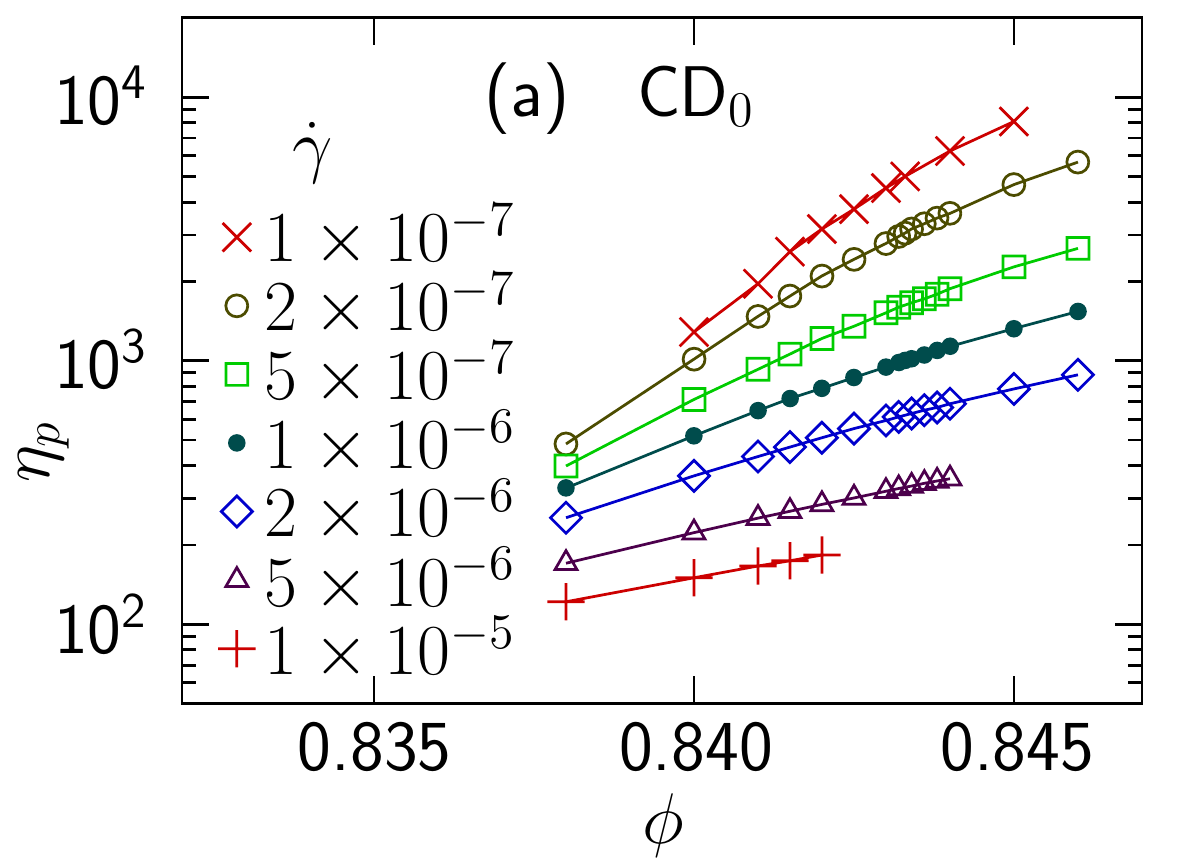}
  \includegraphics[width=1.65in]{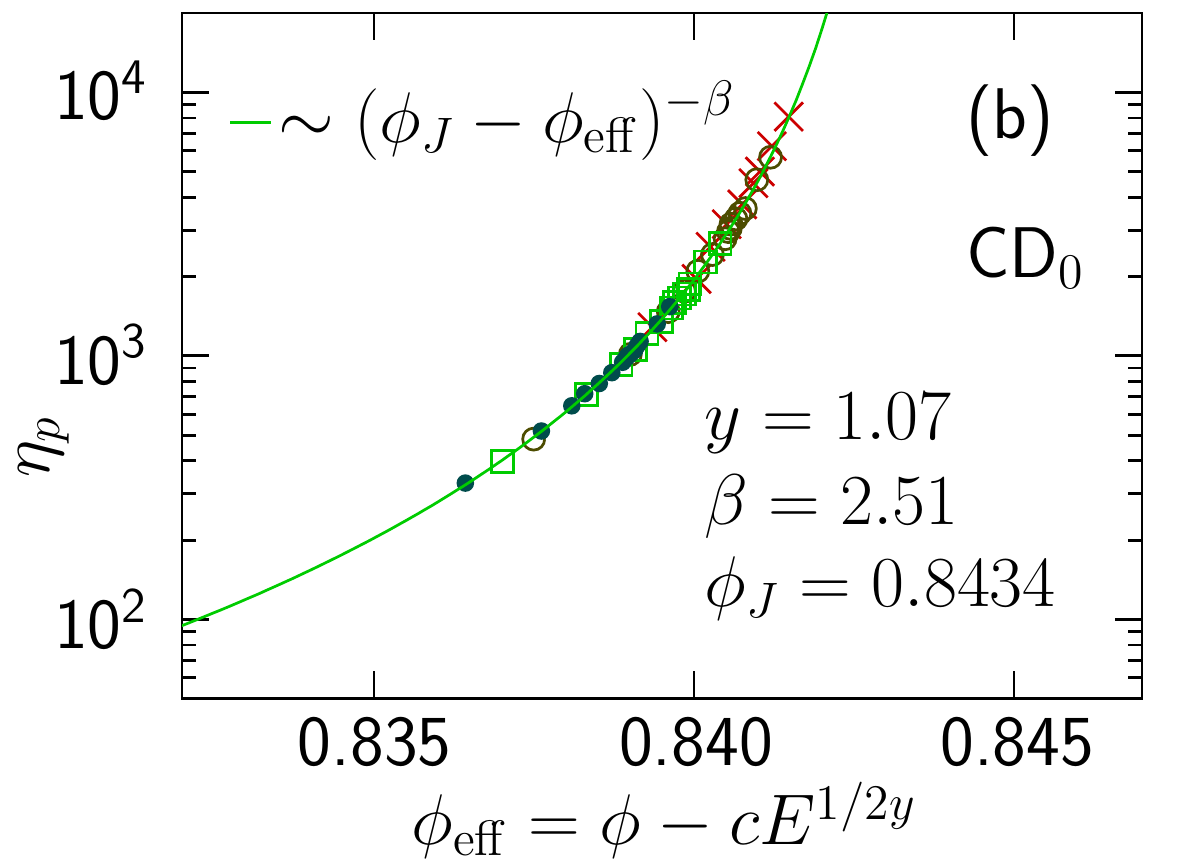}

  \caption{Pressure analog of viscosity $\eta_p\equiv p/\gdot$ for model CD$_0$. Panel (a)
    shows the raw data for $\gdot= 10^{-7}$ through $10^{-5}$ in a narrow interval of
    $\phi$ about $\phi_J$. Solid lines interpolate between the data points.  Panel (b)
    shows our scaling collapse of $\eta_p$ vs.\ $\phieff$, with $\phi_J$, $\beta$ and the
    parameters in \Eqs{phieff}{etaphd} determined from the analysis of data with
    $\gdot\leq 10^{-6}$.  The solid line is the fitted power law scaling function.}
  \label{fig:etap-phi}
\end{figure}

In \Fig{etap-phi} we show the results of such an analysis for CD$_0$.  Panel (a) shows our
raw data for $\eta_p$ vs.\ $\phi$, for several different $\gdot$, in the narrow density
interval around $\phi_J$ that is used for the analysis.  Panel (b) shows the result from
fitting $\eta_p$ for $\gdot\leq10^{-6}$ to \Eq{etaphd}. Our fitted values $\phi_J=0.8434$,
$\beta=2.5\pm0.2$, $y=1.07\pm0.05$ for \CD\ are all very close to our earlier results for
\RD\ ($\phi_J=0.8433$, $\beta=2.58\pm0.10$, $y=1.09\pm0.01$) \cite{Olsson_Teitel:jam-HB}
thus suggesting that the critical behavior in \CD\ is the same as in RD$_0$.

\begin{figure}
    \includegraphics[width=1.8in]{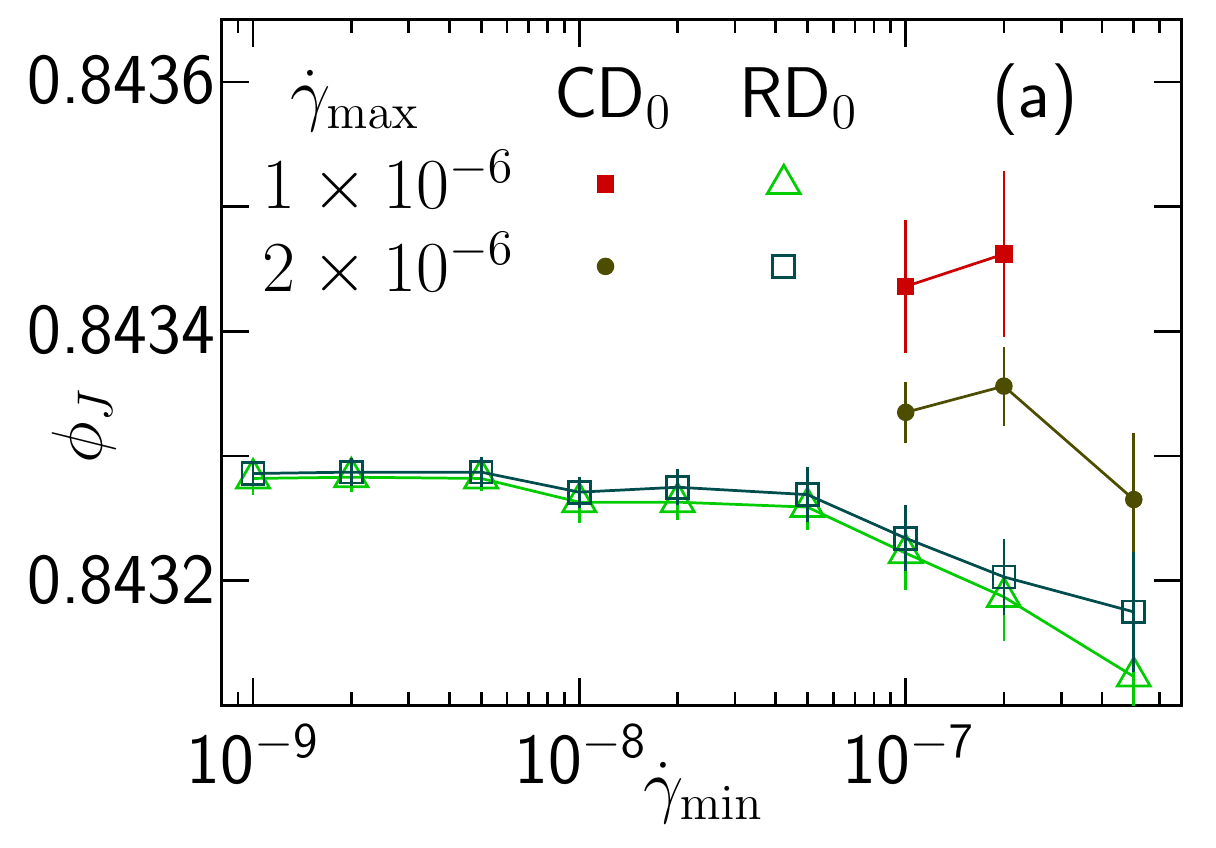}
  \includegraphics[width=1.8in]{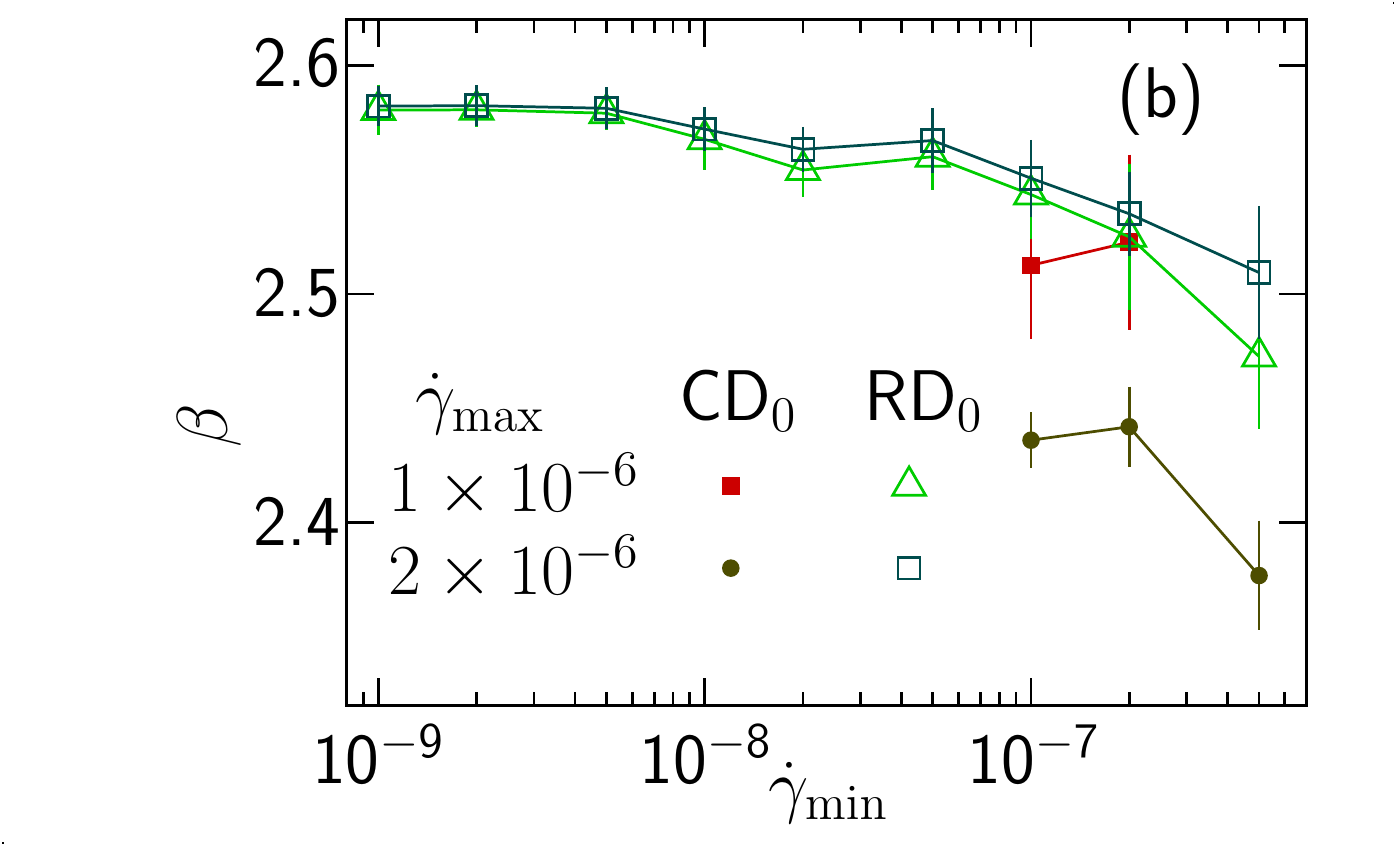}
  \caption{Comparison of critical parameters in models \RD\ and \CD\ for similar ranges of
    strain rate $\gdot$.  Panel (a) shows the jamming packing fraction $\phi_J$ and panel
    (b) the viscosity exponent $\beta$, that result from fits of our data to
    \Eqs{phieff}{etaphd}, for different ranges of $\gdot_\mathrm{min}\le
    \gdot\le\gdot_\mathrm{max}$ and $0.838\leq\phi\leq0.846$. The error bars in the figure
    represent only the statistical errors from the fitting procedure; quoted errors in the
    text include rough estimates of systematic errors, such as arise when varying the
    window in $\phi$ of the data utilized in the fit. We plot results vs.\ varying
    $\gdot_\mathrm{min}$ for two different fixed values of $\gdot_\mathrm{max}=1\times
    10^{-6}$ and $2\times 10^{-6}$.  The open symbols are results for \RD\ and the closed
    symbols are for CD$_0$. For \RD\ we have data down to $\gdot_\mathrm{min}=10^{-9}$,
    however for \CD\ our data goes down to only $\gdot_\mathrm{min}=10^{-7}$.}
      \label{fig:Critical}
\end{figure}

In quoting the fitted values of $\phi_J$, $\beta$ and $y$ we note that our results for
\RD\ include data to much lower strain rates $10^{-9}\le\gdot$ as compared to \CD\ where
$10^{-7}\le\gdot$.  For a more accurate comparison of the two models, we should fit our
data over the same range of strain rates $\gdot$.  We therefore carry out a fitting to
\Eqs{phieff}{etaphd} using data in the interval
$\gdot_\mathrm{min}\le\gdot\le\gdot_\mathrm{max}$.  In \Fig{Critical} we show our results
for $\phi_J$ and $\beta$, where we plot the fitted values vs.\ $\gdot_\mathrm{min}$ for two
different fixed values of $\gdot_\mathrm{max}=1\times 10^{-6}$ and $2\times 10^{-6}$.  For
\RD\ we can extend this procedure down to $\gdot_\mathrm{min}=10^{-9}$, while for \CD\ we
are limited to $\gdot_\mathrm{min}=10^{-7}$.  We see that for equivalent ranges of
$\gdot$, the fitted values of $\beta$ agree nicely for the two models, for the smaller
value of $\gdot_\mathrm{max}$.  We see that $\phi_J$ for \CD\ is just slightly higher than
for RD$_0$.  We cannot say whether this is a systematically significant difference, or
whether $\phi_J$ would decrease slightly to match the value found for \RD\ were we able to
study \CD\ down to comparably small $\gdot$.  Whether or not the $\phi_J$ of the two
models are equal, or just slightly different, the equality of the exponents $\beta$
strongly argues that models \RD\ and \CD\ are in the same critical universality class.

To return to the results of Tighe et al.\ \cite{Tighe_WRvSvH}, we note that their value
$\phi_J=0.8423$ for \CD\ is clearly different from our above value of $0.8434$.  We
believe that this difference is due to two main effects: (i) their data is restricted to
$10^{-5}\le\gdot$ and so does not probe as close to the critical point as we do here, and
(ii) their analysis was based on the scaling of shear viscosity $\eta\equiv\sigma/\gdot$
rather than the pressure viscosity $\eta_p$.  As we have noted previously
\cite{Olsson_Teitel:gdot-scale} corrections to scaling for $\sigma$ are significantly
larger than they are for $p$, and without taking these corrections into account, one
generally finds a lower value for $\phi_J$, such as was also found in the original scaling
analysis of \RD\ \cite{Olsson_Teitel:jamming}.  Their lower value of $\phi_J$, and their
higher window of strain rates $\gdot$, we believe are also responsible for the different
value they find for the exponent describing non-linear rheology exactly at $\phi_J$,
$\sigma\sim p\sim \gdot^q$; they claim $q=1/2$ whereas our present result finds a clearly
different value $q=y/(\beta+y)\approx 0.30$ \cite{Olsson_Teitel:jam-HB}.

\emph{Macroscopic friction}: Finally we consider the macroscopic friction,
$\mu\equiv\sigma/p$.  In Fig.~4 we plot $\mu$ vs $\phi$ for several different values of
strain rate $\gdot$.  We also show results from quasistatic simulations
\cite{Vagberg_VMOT:jam-fss,Vagberg_OT:protocol}, representing the $\gdot\to 0$
limit\cite{quasistatic}.  We use a system with $N=1024$ particles to more explicitly
compare with the results of Tighe et al., who used a similar size system.  While $\mu$ for
the models \RD\ and \CD\ differ slightly at the lower $\phi$, we see that near $\phi_J$
they become essentially equal at the smaller $\gdot$, and both \RD\ and \CD\ approach the
quasistatic limit as $\gdot\to 0$.  We thus conclude that $\mu$ is finite as $\phi$ passes
through $\phi_J$, consistent with recent experiments on foams by Lespiat et al.\
\cite{Lespiat}.

From our fit to $\eta_p$ in Fig.~\ref{fig:etap-phi} we conclude that for both models \CD\
and \RD\ the pressure along the yield stress line, i.e. $\gdot\to 0$, $\phi>\phi_J$,
vanishes upon approaching $\phi_J$ as $p_0\sim (\phi-\phi_J)^y$ with $y\simeq 1.08$.  Our
results in Fig.~4 then argue that the shear stress along the yield stress line,
$\sigma_0$, vanishes similarly, so that $\mu$ stays finite.  However the prediction of
Tighe et al.\ is that the yield shear stress vanishes as $\sigma_0 \sim
(\phi-\phi_J)^{3/2}$.  Were this conclusion correct, we would expect $\mu\sim
(\phi-\phi_J)^{0.42}$, \emph{vanishing} as $\phi\to\phi_J$ from above.  Nothing in Fig. 4,
where we see that $\mu=\sigma/p$ is a monotonically \emph{increasing} function as $\phi$
\emph{decreases}, suggests any such vanishing of $\mu(\phi_J)$.  We thus conclude from
Fig.~4 that the predicted scaling of $\sigma_0$ by Tighe et al.\ is not correct, and
moreover the two models \CD\ and \RD\ behave qualitatively the same for both pressure $p$
and shear stress $\sigma$.

\begin{figure}[h!]
  \includegraphics[width=3.in]{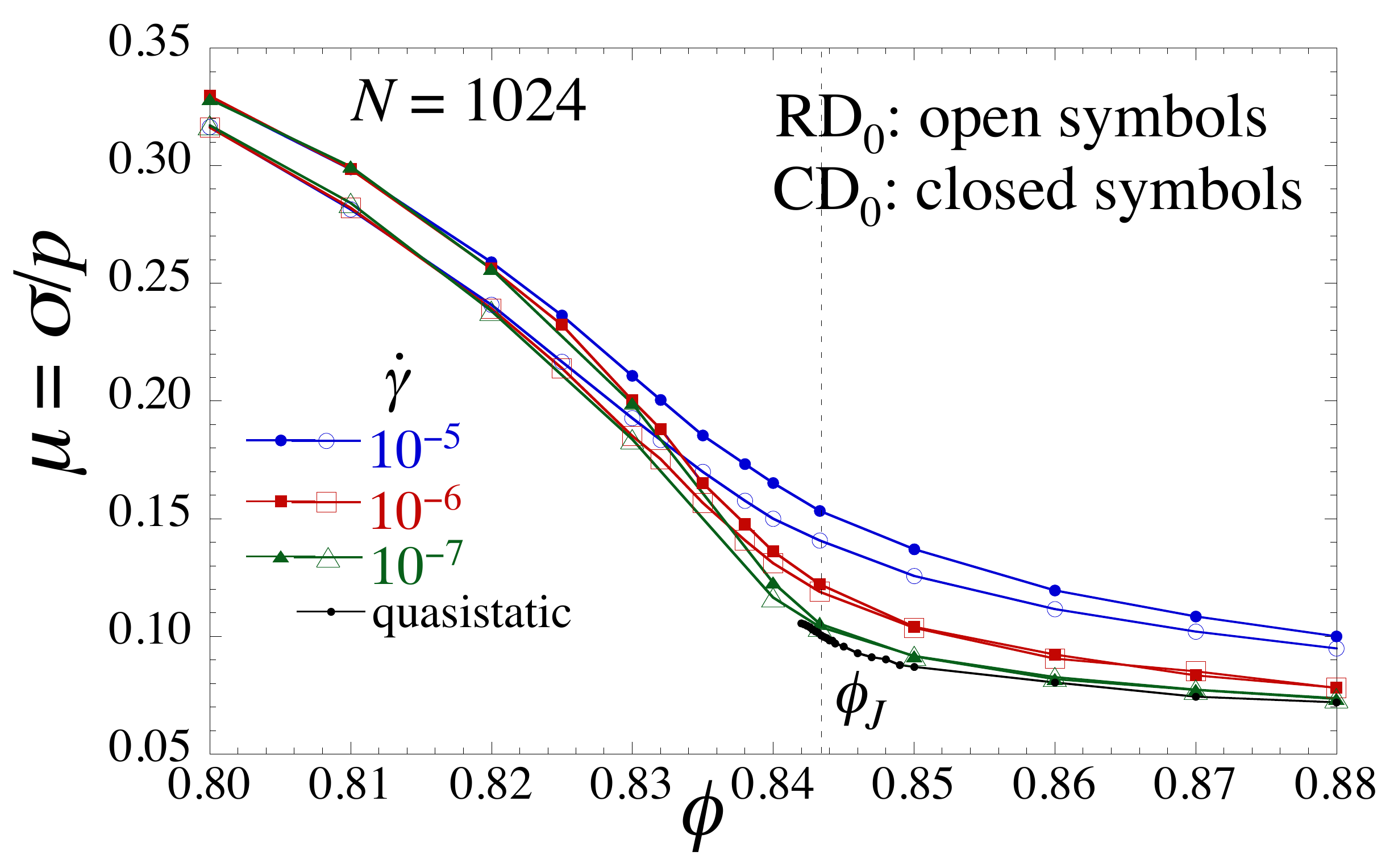}
  \caption{Macroscopic friction $\sigma/p$ vs $\phi$ for different $\gdot$ for models \RD\
    (open symbols) and \CD\ (closed symbols), for a system with $N=1024$ particles.  Also
    shown are results from quasistatic simulations representing the $\gdot\to 0$ limit.}
  \label{f4}
\end{figure}

To conclude, we have examined the issue of the universality of the jamming transition for
overdamped shear-driven frictionless soft-core disks in 2D.  We have considered two
different dissipative models that have been widely used in the literature: the collisional
Durian bubble model \CD\ and its mean field approximation RD$_0$. Contrary to previous
claims \cite{Tighe_WRvSvH} we find clear evidence that \CD\ does exhibit a growing
macroscopically large length $\xi$ that appears to diverge as the jamming critical point
is approached.  We further provide strong evidence that \CD\ and \RD\ are in fact in the
same universality class with the same critical exponents at jamming, and have
qualitatively the same rheological behavior more generally.

We thank A.~J.~Liu, B.~Tighe, M.~van Hecke and M. Wyart for helpful discussions.  This work was
supported by NSF Grant No.\ DMR-1205800 and the Swedish Research Council Grant
No.\ 2010-3725. Simulations were performed on resources provided by the Swedish
National Infrastructure for Computing (SNIC) at PDC and HPC2N.

\newpage
\section{Supplemental Material}
\setcounter{figure}{0}
\setcounter{equation}{0}
\renewcommand{\theequation}{S\arabic{equation}}
\renewcommand{\thefigure}{S\arabic{figure}}
\subsection{Transverse Velocity Correlation Function}

\begin{figure}[h!]
  \includegraphics[width=1.65in]{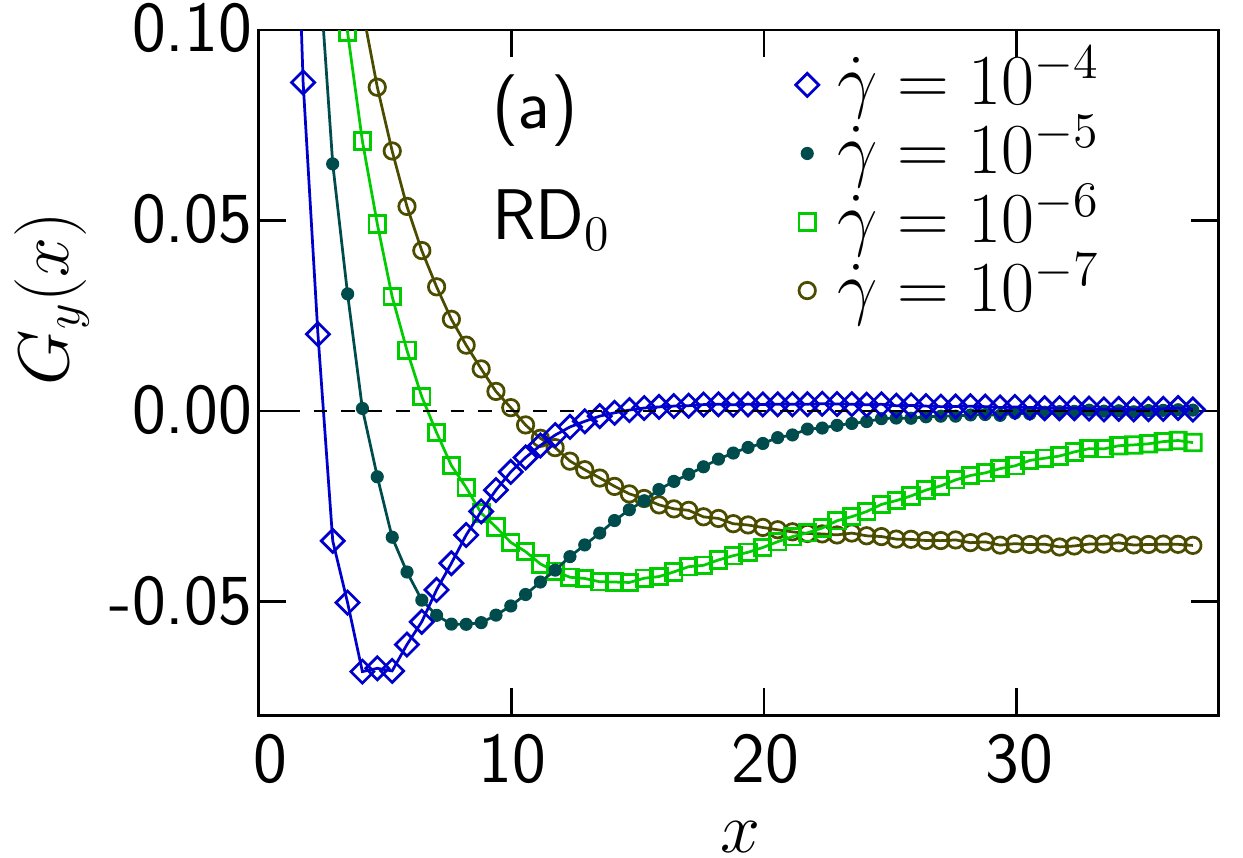}
  \includegraphics[width=1.65in]{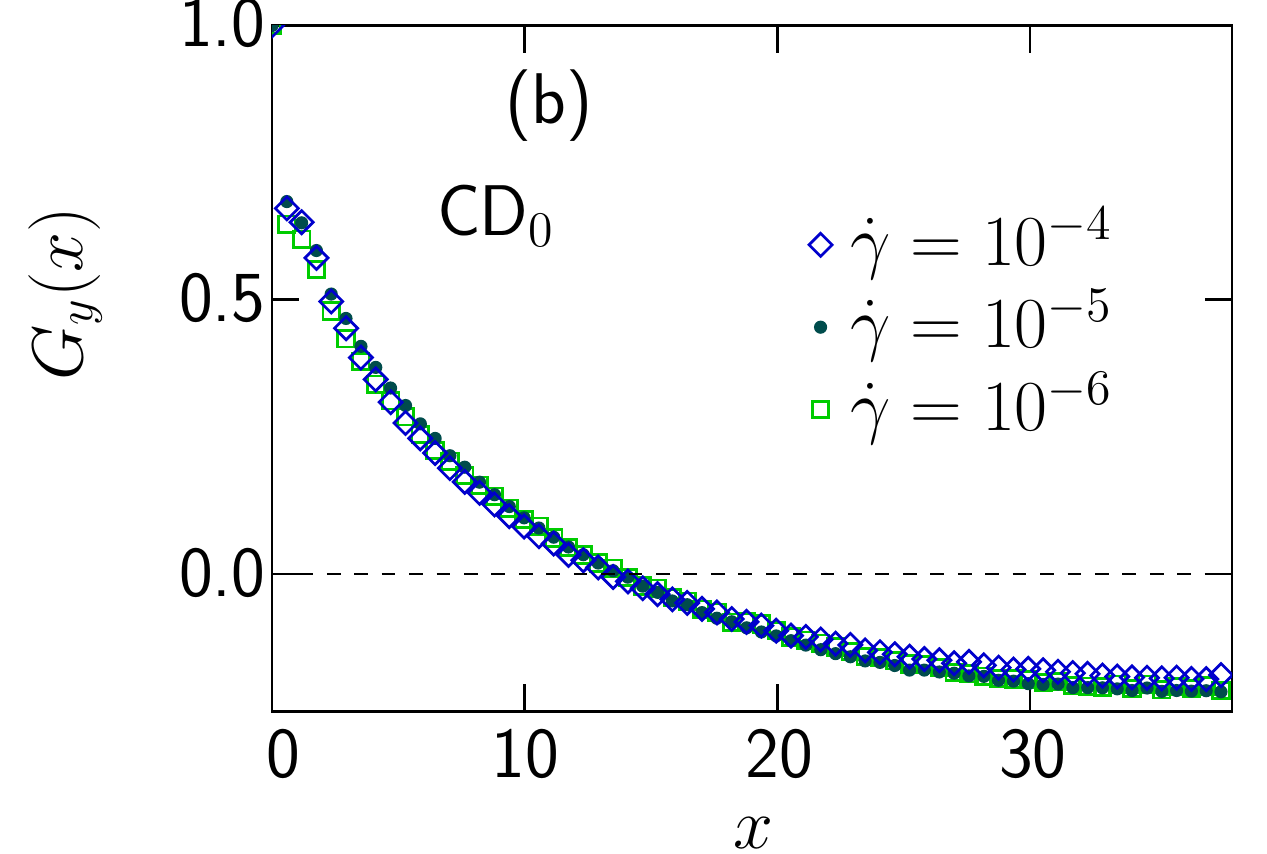}
  \caption{Normalized transverse velocity correlation function $G_y(x)=g_y(x)/g_y(0)$ at
    $\phi=0.8433 \approx\phi_J$ for a system of $N=4096$ particles. Panel (a) is for model \RD\ with shear rates
    $\gdot=10^{-7}$ through $10^{-4}$. Panel (b) for model \CD\ at shear rates $\gdot=10^{-6}$, through $10^{-4}$. }
  \label{f1SM}
\end{figure}

The one quantity for which models \RD\ and \CD\ are clearly different is the transverse
velocity correlation function, $g_y(x)\equiv\langle v_y(0)v_y(x)\rangle$.  Defining the
normalized correlation, $G_y(x)\equiv g_y(x)/g_y(0)$, we plot in Fig.~\ref{f1SM}(a)
$G_y(x)$ vs $x$, for several different values of strain rate $\gdot$, for model \RD\ at
$\phi=0.8433\approx \phi_J$ in a system of $N=4096$ particles.  We see that $G_y(x)$ has a
clear minimum at a distance $x=\ell$, and that $\ell$ increases as $\gdot\to 0$ and one
approaches the critical point.  In Ref.~[\onlinecite{Olsson_Teitel:jamming}] $\ell$ was
interpreted as the diverging correlation length $\xi$.  In \CD\, however, it was found
\cite{Tighe_WRvSvH} that $G_y(x)$ decreases monotonically without any obvious strong
dependence on either $\phi$ or $\gdot$.  In Fig.~\ref{f1SM}(b) we plot $G_y(x)$ vs $x$,
for several different $\gdot$, at $\phi=0.8433\approx \phi_J$ in a system of $N=4096$
particles, confirming this result.

\begin{figure}
\includegraphics[ width=1.6in]{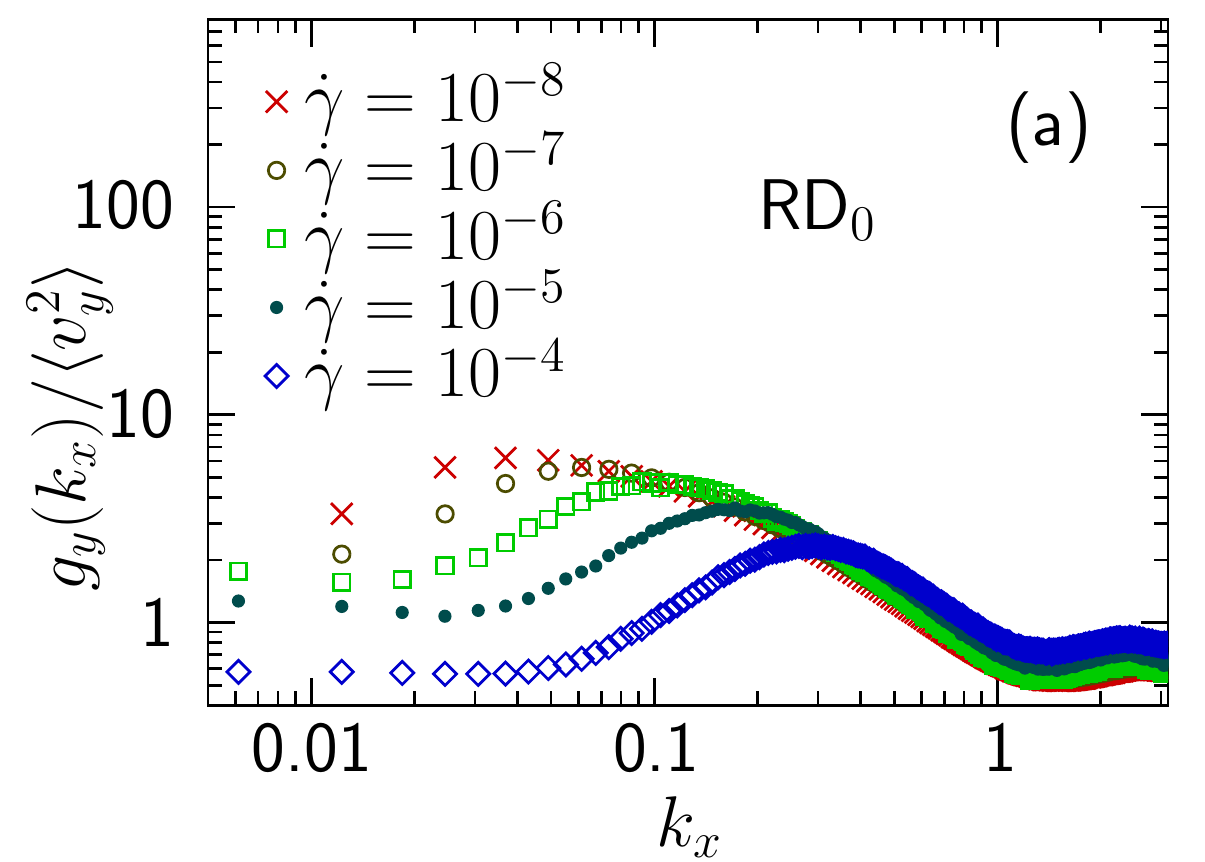}
\includegraphics[width=1.6in]{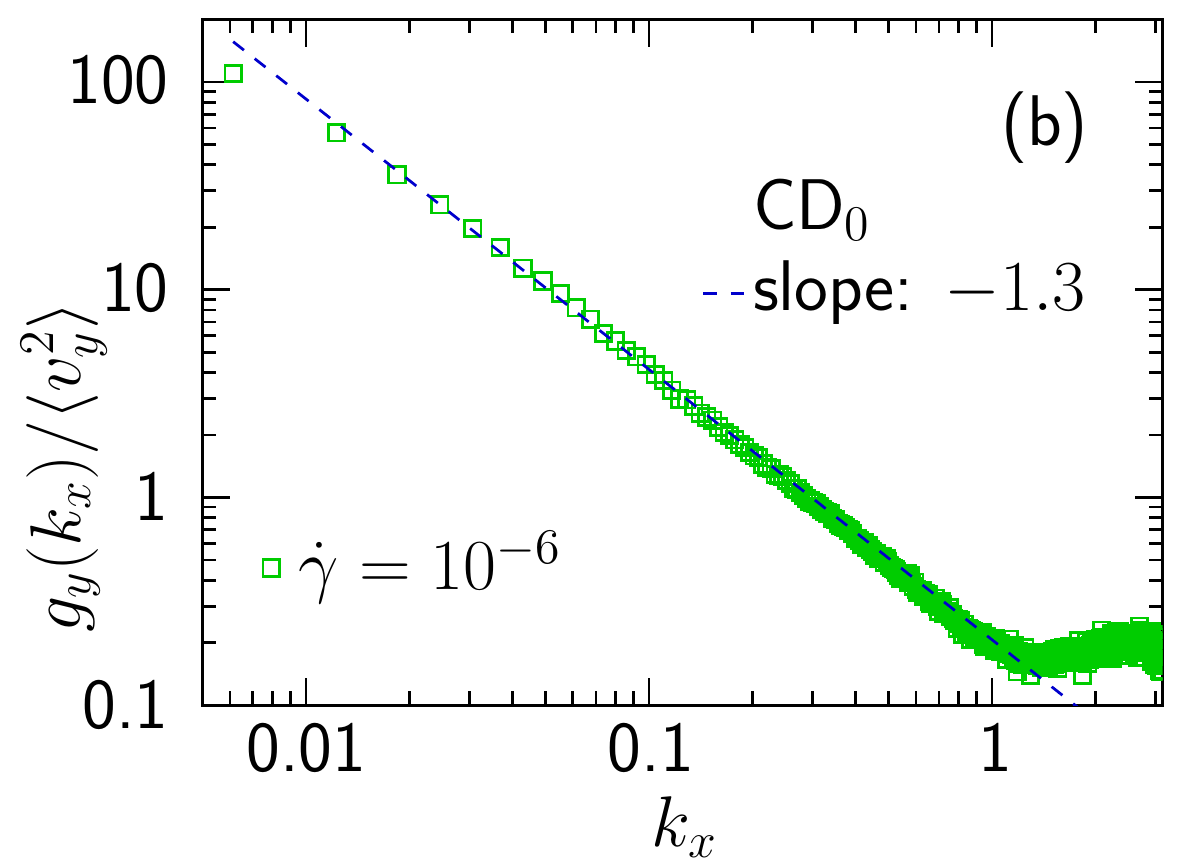}
  \caption{Fourier transform of the transverse velocity correlation function $g_y(k_x)$ at
    $\phi=0.8433 \approx\phi_J$. Panel (a) is for model \RD\ with shear rates
    $\gdot=10^{-8}$ through $10^{-5}$. The peak in $g_y(k_x)$, moving to smaller $k_x$ as
    $\gdot$ decreases, is related to the minimum in the real space $g_y(x)$ moving to
    larger $x$. The algebraic behavior in panel (b) for model \CD\ at $\gdot=10^{-6}$, is
    consistent with the absence of any apparent length scale, as reported in
    \Ref{Tighe_WRvSvH}. The number of particles in these figures are $N=262144$ except for
    the two smallest shear rates for RD$_0$ for which $N=65536$.}
  \label{fig:gy-kx}
\end{figure}

As an alternative way to consider the difference in this correlation between the two models, we now
consider the Fourier transformed correlation $g_y(k_x)=\int dx\, g_y(x) \mathrm{e}^{ik_x x}$, which we
show in \Figs{gy-kx}(a) and \ref{fig:gy-kx}(b) for \RD\ and \CD\ respectively at packing
fraction $\phi=0.8433\approx\phi_J$. 
For \RD\ we see a maximum in $g_y(k_x)$ at a $k_x^*$ that
decreases for decreasing $\dot\gamma$; $\ell\sim1/k_x^*$ gives the corresponding minimum
of the real-space correlation.   For \CD\ we show results only for the single strain rate $\gdot=10^{-6}$ since
from Fig.~\ref{f1SM}(a) we expect no observable difference as $\gdot$ varies.  
We see an algebraic divergence $g_y(k_x)\sim
k_x^{-1.3}$ as $k_x\to 0$. It is this algebraic divergence that causes the real space $G_y(x)$ in \CD\ to become solely a function of $x/L$ as the system length $L$ increases, as was reported in \Ref{Tighe_WRvSvH}.

To try and give a qualitative understanding of this differing behavior of $g_y(k_x)$, we can consider how energy is dissipated in each model.  In RD$_0$ the dissipation is $(1/N)\sum_i \langle|\delta{\bf v}_i|^2\rangle \approx \int d{\bf k}\langle \delta{\bf v}({\bf k})\cdot\delta{\bf v}(-{\bf k})\rangle$.  For CD$_0$, however, the dissipation is $(1/N)\sum_{i,j} \langle|{\bf v}_i-{\bf v}_j|^2\rangle \approx \int d{\bf k}\langle \delta{\bf v}({\bf k})\cdot\delta{\bf v}(-{\bf k})\rangle|{\bf k}|^2$, where the sum is over only neighboring particles $i,j$ in contact. Here $\delta{\bf v}$ is the non-affine part of the particle velocity.  
If we make an equipartition-like ansatz, and assume that as $k\to 0$ all modes ${\bf k}$, and both spatial directions $x$, $y$, contribute equally to the dissipation, we would then conclude that for RD$_0$ $\langle {v}_y({\bf k}){ v}_y(-{\bf k})\rangle\propto{\rm constant}$, while for CD$_0$ $\langle {v}_y({\bf k}){v}_y(-{\bf k})\rangle\propto 1/k^2$.  
Noting that $g_y(k_x)=\int dk_y \langle {v}_y({\bf k}){v}_y(-{\bf k})\rangle$, we then conclude that for \RD\
we have $g(k_x)\propto{\rm constant}$ as $k_x\to 0$, while for CD$_0$ we have the divergence $g(k_x)\propto 1/k_x$.  This saturation of $g_y(k_x)$ for RD$_0$, as compared to the algebraic divergence of $g_y(k_x)$ for CD$_0$, is what is qualitatively seen in Fig.~\ref{fig:gy-kx}.  

The physical reason for this dramatic difference can be viewed as follows.
For CD$_0$, since the dissipation depends only on velocity differences, uniform
translation of a large cluster of particles with respect to the ensemble average flow has
little cost, thus enabling long wavelength fluctuations.  For RD$_0$ the dissipation is
with respect to a fixed background, so uniform translation of a large cluster causes
dissipation that scales with the cluster size; long wavelength fluctuations are
suppressed.  

That the observed divergence in \CD\ is $\sim k_x^{-1.3}$ rather than the simple $k_x^{-1}$ predicted above, suggests that our equipartition ansatz is not quite correct, and that the different modes interact in a non-trivial way.  That the exponent of this divergence is not an integer or simple rational fraction suggests the signature of underlying critical fluctuations, even though the correlation $g_y(x)$ itself does not yield any obvious diverging length
scale.

\subsection{Finite-Size-Scaling of Pressure}

\begin{figure}[h!]
\includegraphics[width=1.65in]{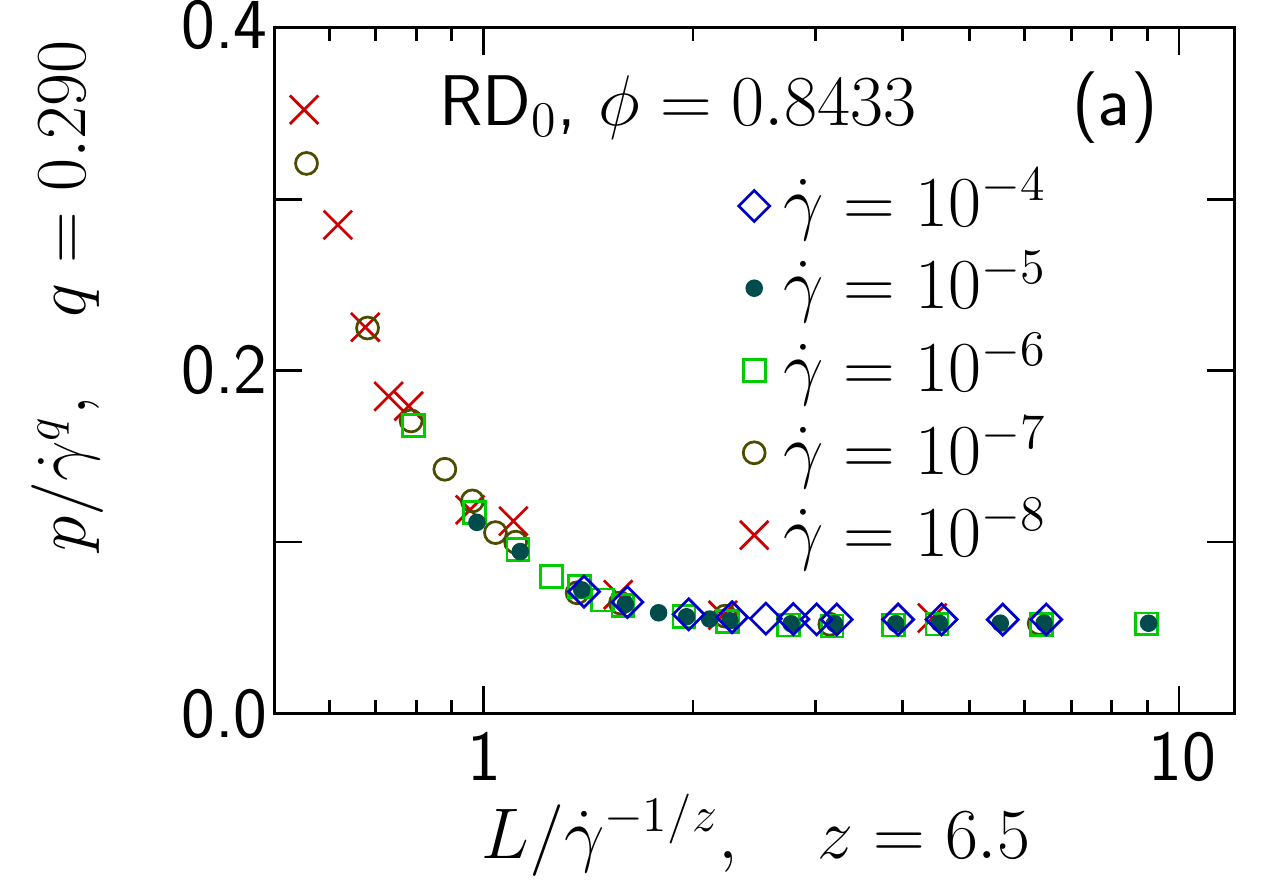}
\includegraphics[width=1.65in]{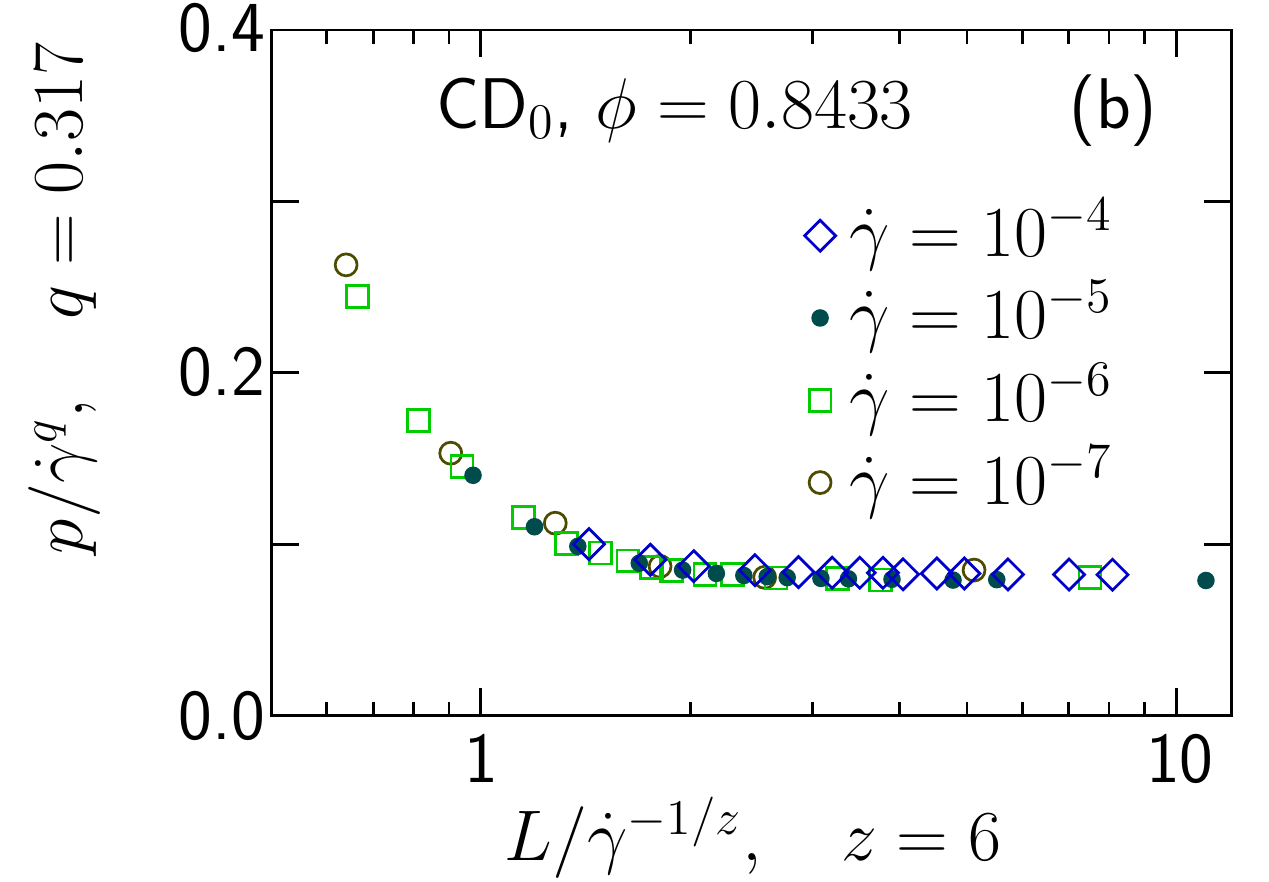}
\caption{Scaling collapse of pressure according to Eq.~(\ref{efss}) for models \RD\ and CD$_0$. }
\label{pffs}
\end{figure}

In Fig.~1 of the main article we showed data for the dependence of pressure $p$ on system
size $L$ at different strain rates $\gdot$, at the jamming fraction $\phi_J\approx
0.8433$.  We argued that these results provided evidence for a similar growing,
macroscopically large, correlation length $\xi$ in both models \RD\ and CR$_0$.  Here we
attempt a finite-size-scaling analysis of this data.  We must note at the outset, however,
that our earlier work \cite{Olsson_Teitel:gdot-scale} demonstrated that it is important to
consider corrections-to-scaling to get accurate values for the exponents at criticality,
and that corrections-to-scaling are in fact large at the smaller sizes $L$ considered in
Fig.~1 of the main article \cite{Vagberg_VMOT:jam-fss}.  Since our data for $p(L)$ is not
extensive enough to try a scaling analysis including corrections-to-scaling, our results
based on a fit to Eq.~(5) must be viewed as providing only {\em effective} exponents
describing the data over the range of parameters considered, rather than the true
exponents asymptotically close to criticality.  We restate Eq.~(5),
\begin{equation}
p(\phi_J,\gdot,L) = L^{-y/\nu}\,{\cal P}(0,\gdot L^z).
\label{efss0}
\end{equation}
We can equivalently write the above in the form
\begin{equation}
p(\phi_J,\gdot,L) =\gdot^{y/z\nu}f(L\gdot^{1/z}),
\label{efss}
\end{equation}
using $f(x)\equiv x^{-y/\nu}\,{\cal P}(0, x^{z})$.  We can now adjust the parameters
$q\equiv y/z\nu$ and $z$ to try and collapse the data to a single common scaling curve.
Plotting $p/\gdot^q$ vs $L\gdot^{1/z}$ we show the results for \RD\ and \CD\ in
Figs.~\ref{pffs}(a) and (b).  For \RD\ we find the effective exponents $z=6.5$ and
$q=0.290$, while for \CD\ we find $z=6.0$ and $q=0.317$. The values of $z$ found in the
present analysis are comparable to the value $z=5.6$ found in the cruder analysis in
Fig.~1(b) of the main article.  Note that for both models the scaling function $f(x)\to$
constant as $x\to \infty$, which gives $p\sim\gdot^q$, $q\equiv y/z\nu$, in the limit of
an infinite sized system.

The closeness of these fitted \emph{effective} exponents for the two models is one more
piece of evidence that \RD\ and \CD\ behave qualitatively the same, and do not have
dramatically different rheology as was claimed by Tighe et al. in
Ref.~\cite{Tighe_WRvSvH}.

Finally we consider how the effective exponents found here compare to the true exponents
asymptotically close to criticality.  From our most accurate analysis
\cite{Olsson_Teitel:gdot-scale} of the critical behavior in RD$_0$, using a large system
size $N=65536$ and including the leading corrections-to-scaling, we have found the
critical exponents $q=y/z\nu=0.28\pm 0.02$ and $y=1.08\pm0.03$, yielding $z\nu=3.9\pm
0.4$.  This value of $q$ is in reasonable agreement with that found above from the
finite-size-scaling analysis of $p(\phi_J,\gdot,L)$.  If we take the value of $z\approx 6$
found in the finite-size-scaling analysis, we would then conclude $\nu\approx 0.65$.  We
note that earlier scaling analyses
\cite{Olsson_Teitel:jamming,OHern_Silbert_Liu_Nagel:2003} that similarly ignored
corrections-to-scaling found similar values for $\nu$.  However our recent
\cite{Vagberg_VMOT:jam-fss} more detailed finite-size-scaling analysis of the correlation
length exponent, which included corrections-to-scaling, found that $\nu\approx 1$,
therefore implying $z\approx 3.9$ as the true critical value.  We thus conclude that the
larger than expected value of $z$ found here from the finite-size-scaling of $p$ is due to
the strong corrections-to-scaling that effect the correlation length at small $L$.

\begin{figure}[h!]
\includegraphics[width=2.5in]{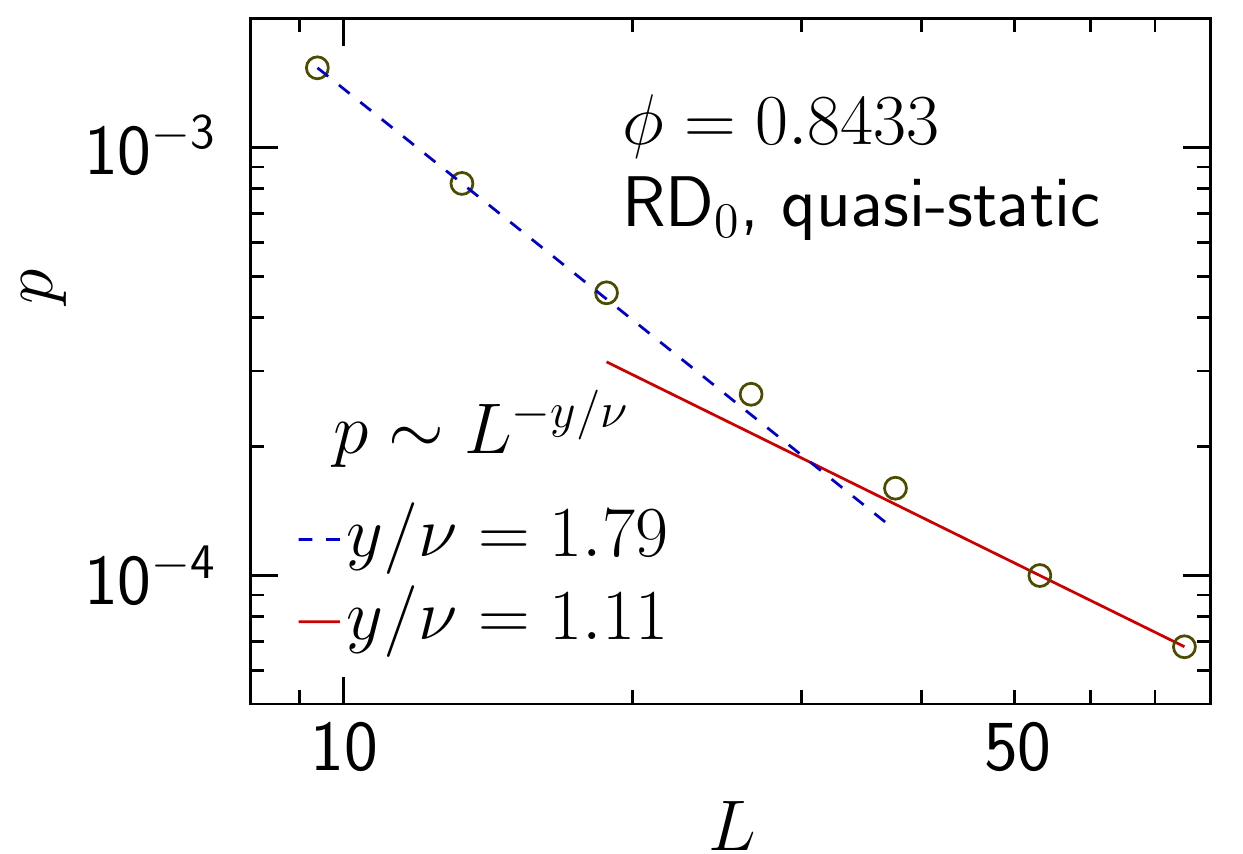}
\caption{Pressure $p$ vs system length $L$ at $\phi_J\approx 0.8433$ for quasistatic
  shearing.  Dashed line is a power law fit to the data at the smallest $L$, giving an
  exponent $y/\nu\approx 1.79$; solid line is a power law fit to the data at the largest
  $L$, giving an exponent $y/\nu\approx 1.11$.}
\label{pqs}
\end{figure}

As another way to see the effect of corrections-to-scaling on the correlation length, in Fig.~\ref{pqs} we plot our results for $p$ vs $L$ at $\phi=0.8433\approx\phi_J$, as obtained from quasistatic simulations \cite{Vagberg_VMOT:jam-fss, Vagberg_OT:protocol} representing the $\gdot\to 0$ limit.
From Eq.~(\ref{efss0}) we expect as $\gdot\to 0$ the behavior, $p\sim L^{-y/\nu}$.  If we fit the data at small $L$ in Fig.~\ref{pqs} to a power law, we then find the exponent, $y/\nu\approx 1.79$.  Using $y=1.08$ this then gives $\nu\approx 0.60$, in rough agreement with the value of $\nu$ obtained from the measured $z$ of our finite-size-scaling of $p$ with $\gdot$.  If, however, we fit the data at only the largest $L$ to a power law, we then find the exponent $y/\nu\approx 1.11$.  Again using $y=1.08$, we then get $\nu\approx 0.97$, in better agreement with the expected $\nu\approx 1$.  Fig.~\ref{pqs} thus shows in a very direct way that corrections-to-scaling are significant for small system lengths $L$.

To conclude this section, although our finite-size-scaling of the pressure data in Fig.~1(a) of the main article is effected by corrections-to-scaling, and so gives a larger value for the dynamic exponent $z$ than we believe is actually the case at criticality, nevertheless the correlation length $\xi$ extracted from this data and shown in Fig.~1(b) demonstrates that \RD\ and \CD\ are behaving qualitatively the same, and that both have a macroscopic length scale $\xi$ that is growing (and we would argue diverging) as the jamming transition is approached.

\subsection{Effect of Finite Mass on Model CD}

\begin{figure}[h!]
\includegraphics[width=1.6in]{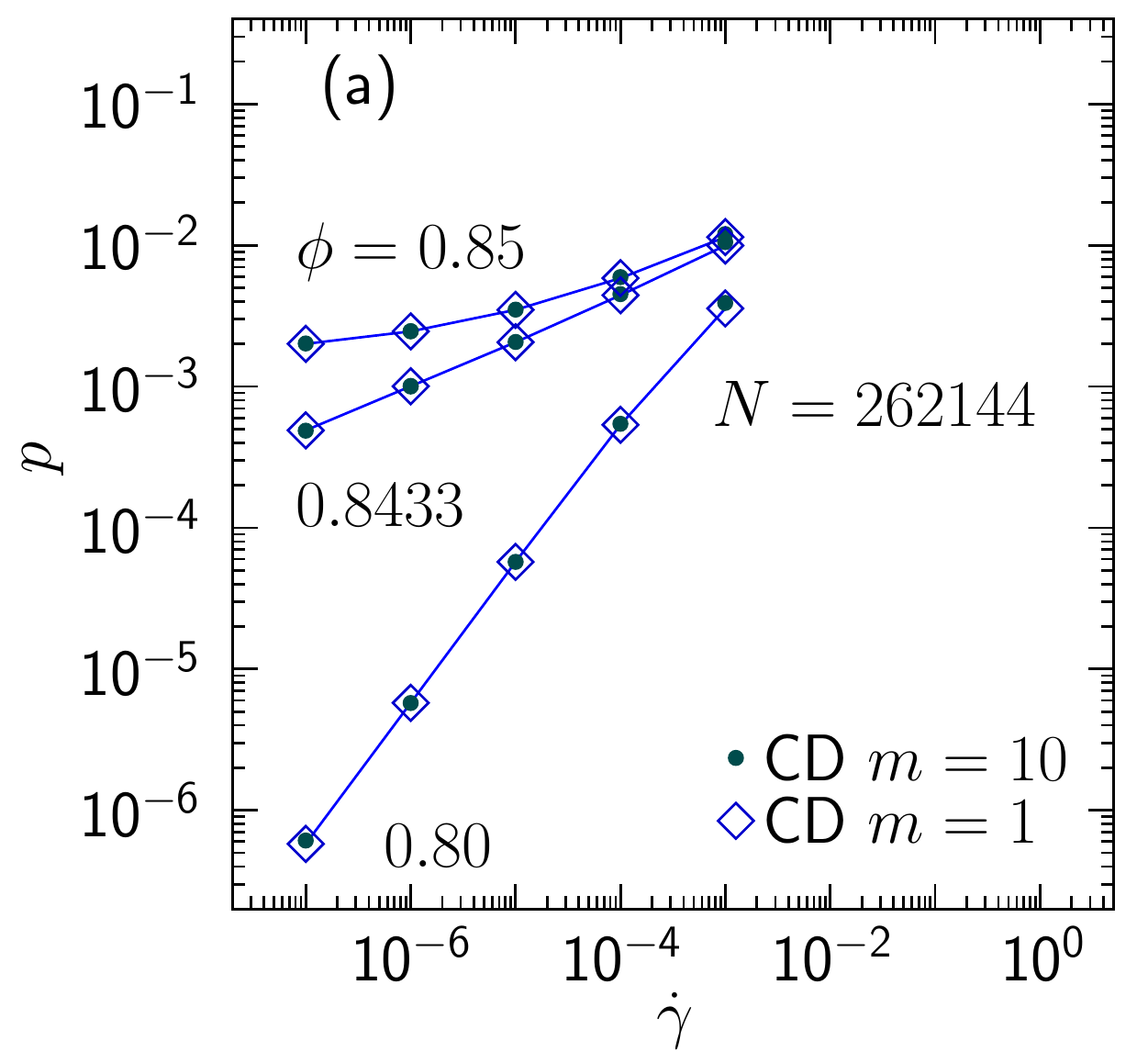}
\includegraphics[width=1.6in]{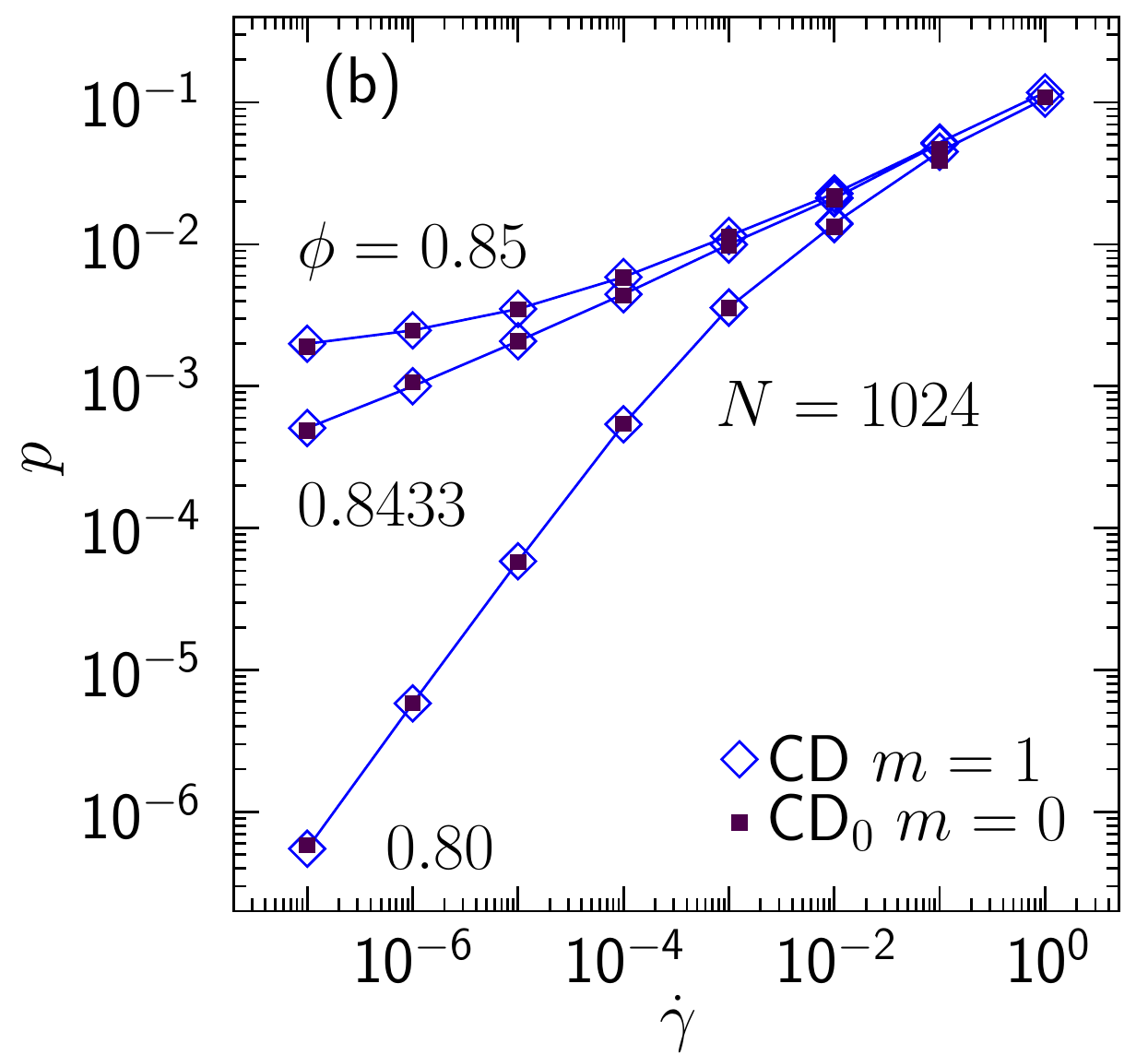}
  \caption{Pressure $p$ vs.\ shear strain rate $\gdot$ at packing fractions $\phi=0.80$,
    0.8433, 0.85 for: (a) model CD with $m=1$ and $m=10$ for $N=262144$ particles, and
    (b) model CD with $m=1$ and model CD$_0$ with $m=0$ for $N=1024$ particles.}
  \label{f5}
\end{figure}

We wish to verify that the mass parameter $m=1$, which we use in model CD, is
indeed sufficiently small so as to put our results in the overdamped $m\to 0$ limit
corresponding to model CD$_0$, for the range of parameters studied here. In
Fig.~\ref{f5}(a) we show results for the elastic part of the pressure $p$ vs $\gdot$ for
model CD, with $N= 262144$ particles, at three different packing fractions: $\phi=0.80$,
$\phi=0.8433\approx\phi_J$, and $\phi=0.85$.  We compare results for two different mass
parameters, $m=1$ and $m=10$.  We see that the results agree perfectly for small $\gdot$;
significant differences are only found for $\gdot\geq10^{-3}$ which is higher than the
largest shear rate used in our scaling analysis. In Fig.~\ref{f5}(b) we similarly compare
results for model CD with $m=1$ with explicit results for model CD$_0$, as obtained from
simulations using the more costly matrix inversion dynamics for $m=0$.  In this case we
are restricted to $N=1024$ particles because our algorithm for CD$_0$ scales as $N^2$.  We
see that in all cases there is no observed difference between the two models.  Thus we
conclude that our results from CD with $m=1$ are indeed in the overdamped $m\to 0$ limit.


\begin{thebibliography}{10}
\bibitem{quasistatic}%
  \BibitemOpen
  \bibinfo {note} {Quasistatic simulations only accurately compute $\sigma/p$
  for jammed states, where $\sigma$ and $p$ are both finite. The contribution
  from flowing, unjammed states, where $\sigma/p=\eta/\eta_p$ (where
  $\eta=\sigma/\gdot$ is the shear viscosity), is not properly accounted for in
  a naive calculation since $\sigma$ and $p$ separately vanish as $\gdot\to 0$
  and so do not contribute to the average. Hence we only show quasistatic
  results for $0.842 \le \phi$, where for $N=1024$ we find roughly half or more
  of the configurations sampled during quasistatic shearing are jammed
  \cite{Vagberg_VMOT:jam-fss}. We note that the model of Tighe et al.\
  \cite{Tighe_WRvSvH} applies only above $\phi_J$, and so is in the region
  where the quasistatic calculation is valid.}%
  \bibAnnoteFile{Stop}{quasistatic}%
\end{thebibliography}
\end{document}